\input{aipcheck}

\documentclass[
    ,final            
  ]
  {aipproc}

\layoutstyle{6x9}

\begin{document}

\title{MN Dra - In-the-Gap Dwarf Nova With Negative Superhumps}


\author{E. Pavlenko}{
  address={Crimean Astrophysical Observatory,  Crimea,  Ukraine}
}
\author{T. Kato}{
  address={Kioto University,  Japan}
}
\author{M. Andreev}{
  address={Terskol Branch  of the RAS Institute of  Astronomy, Russia}
}
\author{A. Sklyanov}{
  address={Kazan State  University, Russia}
}
\author{A. Zubareva}{
  address={RAS Institute of Astronomy,  Russia}
}
\author{D. Samsonov}{
  address={Crimean Astrophysical Observatory, Crimea, Ukraine}
}
\author{I. Voloshina}{
  address={Sternberg Astronomical Institute, Russia}
}
\author{V. Metlov}{
  address={Sternberg Astronomical Institute, Russia}
}
\author{S. Shugarov}{
  address={Sternberg Astronomical Institute, Russia}
}
\author{N. Parakhin}{
  address={International Centre for Astronomical, Medical and Ecological Research, Ukraine}
}
\author{A. Golovin}{
  address={Main Astronomical Observatory of the National Academy of Sciences, Ukraine}
}
\author{O. Antoniuk}{
  address={Crimean Astrophysical Observatory, Crimea, Ukraine}
  }
\begin{abstract}
 The multi-site photometric observations of MN Dra were made over
77 nights in August-November, 2009. The total exposure was 433
hours. During this time the binary underwent two superoutbursts
and five normal outbursts. During the course of first
superoutburst  period of positive superhumps decreased with
extremely large $\dot P = -1.5 \times 1.0^{-4}$ for SU UMa-like
dwarf novae, confirming  known behavior of MN Dra [1]. Between the
superoutbursts MN Dra displayed  negative superhumps. Their period
changed cyclically around 0.096-day value.

\end{abstract}

\maketitle

\section{introduction}
MN Dra is an SU UMa-type dwarf nova in the period gap [2] with a
short supercycle of $\sim60$ day [3]. During the superoutbursts it
displays the positive
  0.105-day superhumps [3]. The period of
  superhumps decreases with  a $\dot P$ of an order
   of $10^{-4}$, extremely large for SU UMa-type dwarf novae [1].
   In 2009 after the end of the
   May superoutburst we first found the large-amplitude brightness
   variations existing both in quiescence and during normal outbursts [4]. These
   variations have period 0.096 day and sometimes reach  the amplitude
   of $1^{m}.4$ when the object is in low quiescence. Although the orbital period is still unknown, we
   interpreted that these variations are negative superhumps caused by the nodal precession
   of a tilted accretion disk.
   In order to study the behavior of the 0.096-day variations  we undertook a new campaign in 2009
   that covered almost two consecutive supercycles.

\section{Observations}

  The CCD photometrical observations of MN Dra were carried out
in the Crimean astrophysical observatory with the Shajn 2.6-m
mirror and the 38-cm Cassegrain telescopes, in the Crimean
laboratory of the Sternberg astronomical institute with the 60-cm
Zeiss telescope, in the Tatranska Lomnica observatory with the
50-cm telescope and in the Terskol observatory with the 60-cm
telescope. The multi-site observations were made over 77 nights
(433 hours) in August-November and covered two superoutbursts and
five normal outbursts. The observations were performed in $R$ band
or in white light. In the later case the data were reduced to $R$.
The overall light curve is shown in Fig.1.

\section{Superoutbursts}
During both superoutbursts MN Dra showed the positive superhumps.
Their period decreased with time. In Fig. 2 the evolution of the
superhump period for the first superoutburst is shown. The
$(O-C)^{'}$s were calculated for the superhump maxima accordingly
the ephemeris

\begin{equation}
HJD_{max}=2455023.28+0.105416\times E \label{ionflux}
\end{equation}

It is seen that period of positive superhumps decreases with $\dot
P = -1.5\times10^{-4}$ during  135 cycles. That confirms the
 extreme value of  $\dot P$ for the previous superoutbursts of MN
 Dra $-1.0\times10^{-4}$ [1]. Nogami et al. [3]
 measured an even more extreme value of $\dot P = -17\times10^{-4}$.

\begin{figure}
  \includegraphics[height=0.21\textheight]{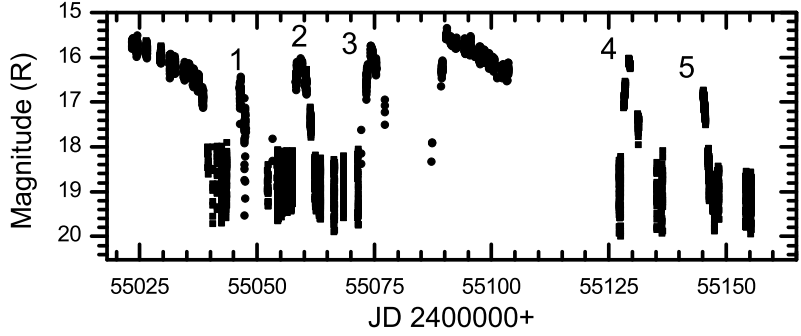}
  \caption{Light curve of MN Dra over two supercycles}
\end{figure}


\begin{figure}
  \includegraphics[height=0.2\textheight]{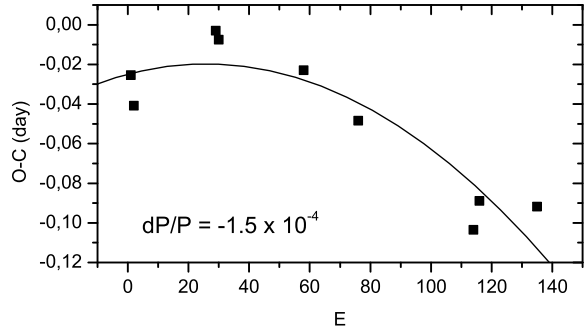}
  \caption{$(O-C)^{'}$s of the positive superhump maxima for the first superoutburst}
\end{figure}

\section{Evolution of negative superhumps}

Outside the superoutburst there are the brightness variations in
minimum and normal outbursts interpreted that these variations are
the negative superhumps. As was shown earlier (Pavlenko et al.,
2010), the amplitude of superhumps varied from $0^{m}.1 - 0^{m}.2$
in the outburst maximum to $1^{m} - 1^{m}.4$ in minimum. The
examples of the light curves in minimum, start and end of the
outburst decline are shown in Fig. 3. They show a growth of
amplitude when brightness decreases. In a scale of intensities the
amplitudes of all the data are situated within the same strip. It
is worth noting that the data in minimum that are well separated
from the normal outbursts display the  dependence on the mean
intensity (see Fig. 4). The times of maxima were determined for
the negative superhumps. We used a period of 0.09598 day for
calculating of the $(O-C)^{'}$s. The $(O-C)^{'}$s during the
supercycle display a cyclic change.  In Fig. 5 the $(O-C)^{'}$s
variations are presented together with light curves of normal
outbursts. One could see that superhump period varies cyclically
between normal outbursts around 0.096-day period, being the
longest  at the end of quiescent state preceding the normal
outburst. These variations look like a jump-like switching of the
period of negative superhumps from the longest value to the
shortest one during the beginning of a normal outburst. Then the
period of negative superhumps smoothly increases to its previous
value.

\begin{figure}
  \includegraphics[height=0.17\textheight]{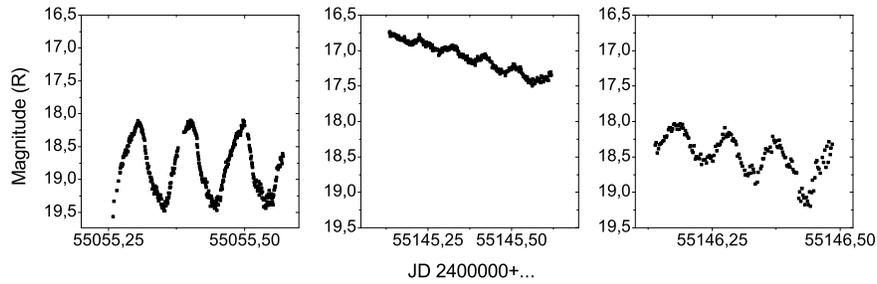}
  \caption{Examples of  original light curves containing the negative superhumps for the quiescent state (left),
  leaving the top of a normal outburst (middle) and approaching to quiescence (right)}
\end{figure}

\begin{figure}
  \includegraphics[height=0.27\textheight]{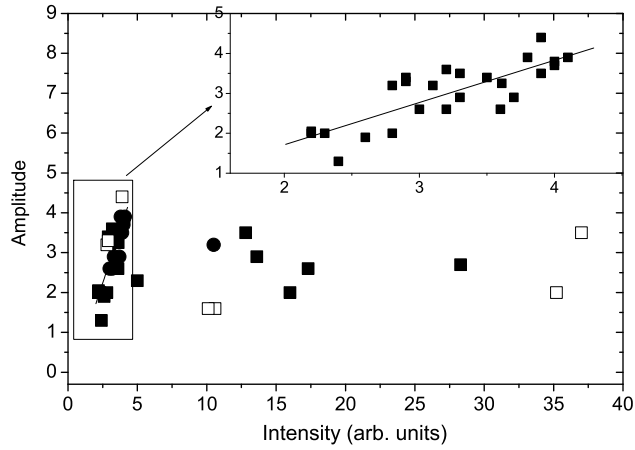}
  \caption{Dependence of the negative superhumps amplitude on the mean intensity in minimum. Dependence
  of amplitude on the mean intensity for the data in quiescence is given in detail in the breakout. The data
  obtained with ZTSh are designed by filled circles, with 60-cm (Terskol) - by filled squares and with
  60-cm (SAI) - by open squares}
\end{figure}

\begin{figure}
  \includegraphics[height=0.38\textheight]{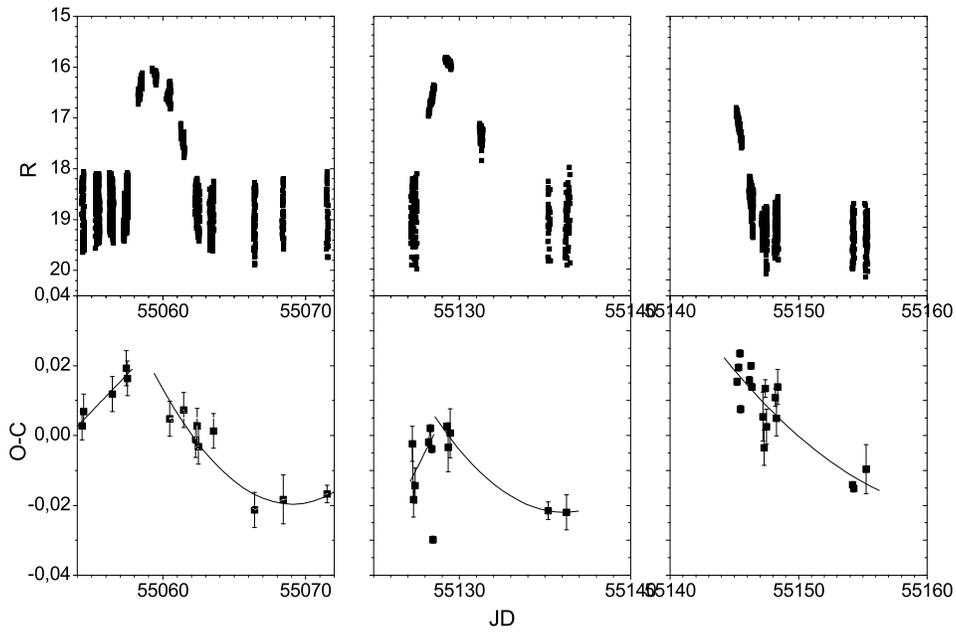}
  \caption{The behavior of the  $(O-C)^{'}$s of negative superhumps (lower part of figure) for three normal outbursts
  (upper part of figure)}
\end{figure}

\section{Conclusion}
We first established the cycling change of the $(O-C)^{'}$s of
negative superhumps  and the dependence of  amplitudes of negative
superhumps on  mean intensity of the system. More data (both
spectroscopic and photometric) are needed to determine the orbital
period of MN Dra and to understand how the outbursts impact on the
nodal precession.

\begin{theacknowledgments}
  E. Pavlenko and A. Golovin are grateful to the LOC for possibility to
  participate in the conference.
  This work was partially supported by grants F 28.2/081 of
  Ukrainian Fund of Fundamental Researches, RFBR 10-02-90458 and VEGA-2/0038/10.
\end{theacknowledgments}



\bibliographystyle{aipproc}   

\bibliography{sample}

\IfFileExists{\jobname.bbl}{}
 {\typeout{}
  \typeout{******************************************}
  \typeout{** Please run "bibtex \jobname" to optain}
  \typeout{** the bibliography and then re-run LaTeX}
  \typeout{** twice to fix the references!}
  \typeout{******************************************}
  \typeout{}
 }

\end{document}